%% file: main.tex
\documentclass{article}

\usepackage{microtype}
\usepackage{graphicx}
\usepackage{subcaption}
\usepackage{booktabs} 
\usepackage{amsmath} 
\usepackage{mathtools} 
\usepackage{amssymb}
\usepackage[frozencache]{minted} 
\DeclareMathOperator*{\argmax}{arg\, max}

\usepackage{hyperref}

\usepackage[accepted]{icml2021}

\icmltitlerunning{Conditional Variational Autoencoder with Adversarial Learning for End-to-End Text-to-Speech}

\begin{document}

\twocolumn[
\icmltitle{Conditional Variational Autoencoder with Adversarial Learning for End-to-End Text-to-Speech}


\icmlsetsymbol{equal}{*}

\begin{icmlauthorlist}
\icmlauthor{Jaehyeon Kim}{kaen}
\icmlauthor{Jungil Kong}{kaen}
\icmlauthor{Juhee Son}{kaen,kaist}
\end{icmlauthorlist}

\icmlaffiliation{kaen}{Kakao Enterprise, Seongnam-si, Gyeonggi-do, Republic of Korea}
\icmlaffiliation{kaist}{School of Computing, KAIST, Daejeon, Republic of Korea}

\icmlcorrespondingauthor{Jaehyeon Kim}{jay.xyz@kakaoenterprise.com}

\icmlkeywords{Machine Learning, ICML}

\vskip 0.3in
]



\printAffiliationsAndNotice{}  

\begin{abstract}

Several recent end-to-end text-to-speech (TTS) models enabling single-stage training and parallel sampling have been proposed, but their sample quality does not match that of two-stage TTS systems. In this work, we present a parallel end-to-end TTS method that generates more natural sounding audio than current two-stage models. Our method adopts variational inference augmented with normalizing flows and an adversarial training process, which improves the expressive power of generative modeling. We also propose a stochastic duration predictor to synthesize speech with diverse rhythms from input text. With the uncertainty modeling over latent variables and the stochastic duration predictor, our method expresses the natural one-to-many relationship in which a text input can be spoken in multiple ways with different pitches and rhythms. A subjective human evaluation (mean opinion score, or MOS) on the LJ Speech, a single speaker dataset, shows that our method outperforms the best publicly available TTS systems and achieves a MOS comparable to ground truth.

\end{abstract}

\input{sections/introduction}
\input{sections/method}
\input{sections/experiments}

\input{sections/results}
\input{sections/relatedwork}

\input{sections/conclusion}
\input{sections/acknowledgements}

\bibliography{main}
\bibliographystyle{icml2021}

\input{sections/appendix}

\end{document}

%% file: sections/introduction.tex
\section{Introduction}

Text-to-speech (TTS) systems synthesize raw speech waveforms from given text through several components. With the rapid development of deep neural networks, TTS system pipelines have been simplified to two-stage generative modeling apart from text preprocessing such as text normalization and phonemization. The first stage is to produce intermediate speech representations such as mel-spectrograms~\cite{shen2018natural} or linguistic features~\cite{oord2016wavenet} from the preprocessed text,\footnote{Although there is a text preprocessing step in TTS systems, We herein use preprocessed text interchangeably with the word ``text".} and the second stage is to generate raw waveforms conditioned on the intermediate representations~\cite{oord2016wavenet, kalchbrenner2018efficient}. Models at each of the two-stage pipelines have been developed independently.

Neural network-based autoregressive TTS systems have shown the capability of synthesizing realistic speech~\cite{shen2018natural,li2019neural}, but their sequential generative process makes it difficult to fully utilize modern parallel processors. To overcome this limitation and improve synthesis speed, several non-autoregressive methods have been proposed. In the text-to-spectrogram generation step, extracting attention maps from pre-trained autoregressive teacher networks~\cite{ren2019fastspeech,peng2020non} is attempted to decrease the difficulty of learning alignments between text and spectrograms. More recently, likelihood-based methods further eliminate the dependency on external aligners by estimating or learning alignments that maximize the likelihood of target mel-spectrograms~\cite{zeng2020aligntts, miao2020flow, kim2020glow}. Meanwhile, generative adversarial networks (GANs)~\cite{goodfellow2014generative} have been explored in second stage models. GAN-based feed-forward networks with multiple discriminators, each distinguishing samples at different scales or periods, achieve high-quality raw waveform synthesis~\cite{kumar2019melgan, binkowski2019high, kong2020hifi}.

Despite the progress of parallel TTS systems, two-stage pipelines remain problematic because they require sequential training or fine-tuning~\cite{shen2018natural,weiss2020wave} for high-quality production wherein latter stage models are trained with the generated samples of earlier stage models. In addition, their dependency on predefined intermediate features precludes applying learned hidden representations to obtain further improvements in performance. Recently, several works, i.e., FastSpeech 2s~\cite{ren2021fastspeech} and EATS~\cite{donahue2021endtoend}, have proposed efficient end-to-end training methods such as training over short audio clips rather than entire waveforms, leveraging a mel-spectrogram decoder to aid text representation learning, and designing a specialized spectrogram loss to relax length-mismatch between target and generated speech. However, despite potentially improving performance by utilizing the learned representations, their synthesis quality lags behind two-stage systems.

In this work, we present a parallel end-to-end TTS method that generates more natural sounding audio than current two-stage models. Using a variational autoencoder (VAE)~\cite{kingma2013auto}, we connect two modules of TTS systems through latent variables to enable efficient end-to-end learning. To improve the expressive power of our method so that high-quality speech waveforms can be synthesized, we apply normalizing flows to our conditional prior distribution and adversarial training on the waveform domain. In addition to generating fine-grained audio, it is important for TTS systems to express the one-to-many relationship in which text input can be spoken in multiple ways with different variations (e.g., pitch and duration). To tackle the one-to-many problem, we also propose a stochastic duration predictor to synthesize speech with diverse rhythms from input text. With the uncertainty modeling over latent variables and the stochastic duration predictor, our method captures speech variations that cannot be represented by text.

Our method obtains more natural sounding speech and higher sampling efficiency than the best publicly available TTS system, Glow-TTS~\cite{kim2020glow} with HiFi-GAN~\cite{kong2020hifi}.
We make both our demo page and source-code publicly available.\footnote{Source-code: \href{https://github.com/jaywalnut310/vits}{https://github.com/jaywalnut310/vits}\\ Demo: \href{https://jaywalnut310.github.io/vits-demo/index.html}{https://jaywalnut310.github.io/vits-demo/index.html}}

%% file: sections/method.tex
\section{Method}
\begin{figure*}[ht]
    \vskip 0.2in
    \begin{center}
        \begin{subfigure}{.62\textwidth}
            \centering
            \includegraphics[width=1.\linewidth]{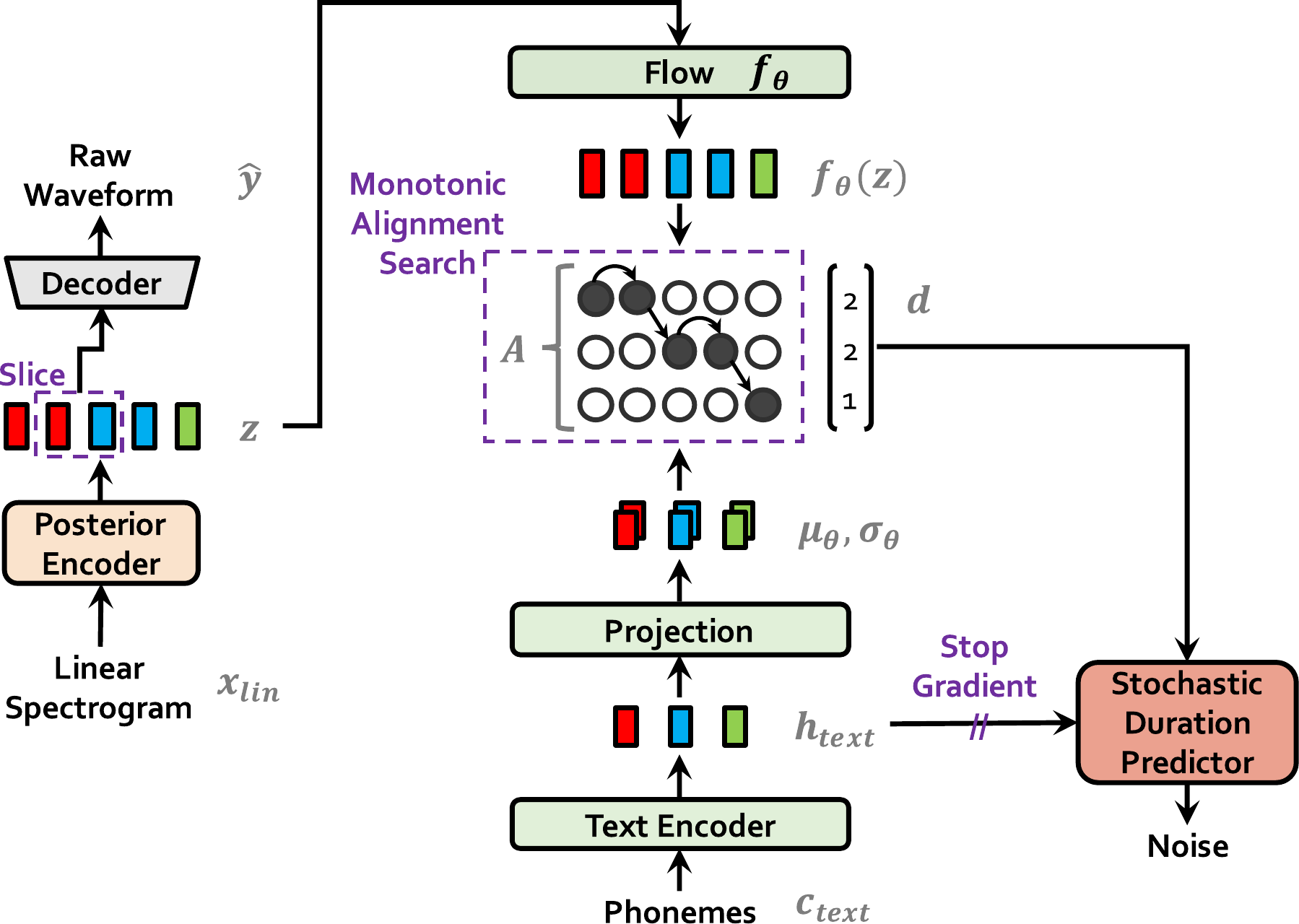}
            \vskip 0.1in
            \caption{Training procedure}
            \label{fig:training}
        \end{subfigure}\hfill%
        \begin{subfigure}{.37\textwidth}
            \centering
            \includegraphics[width=1.\linewidth]{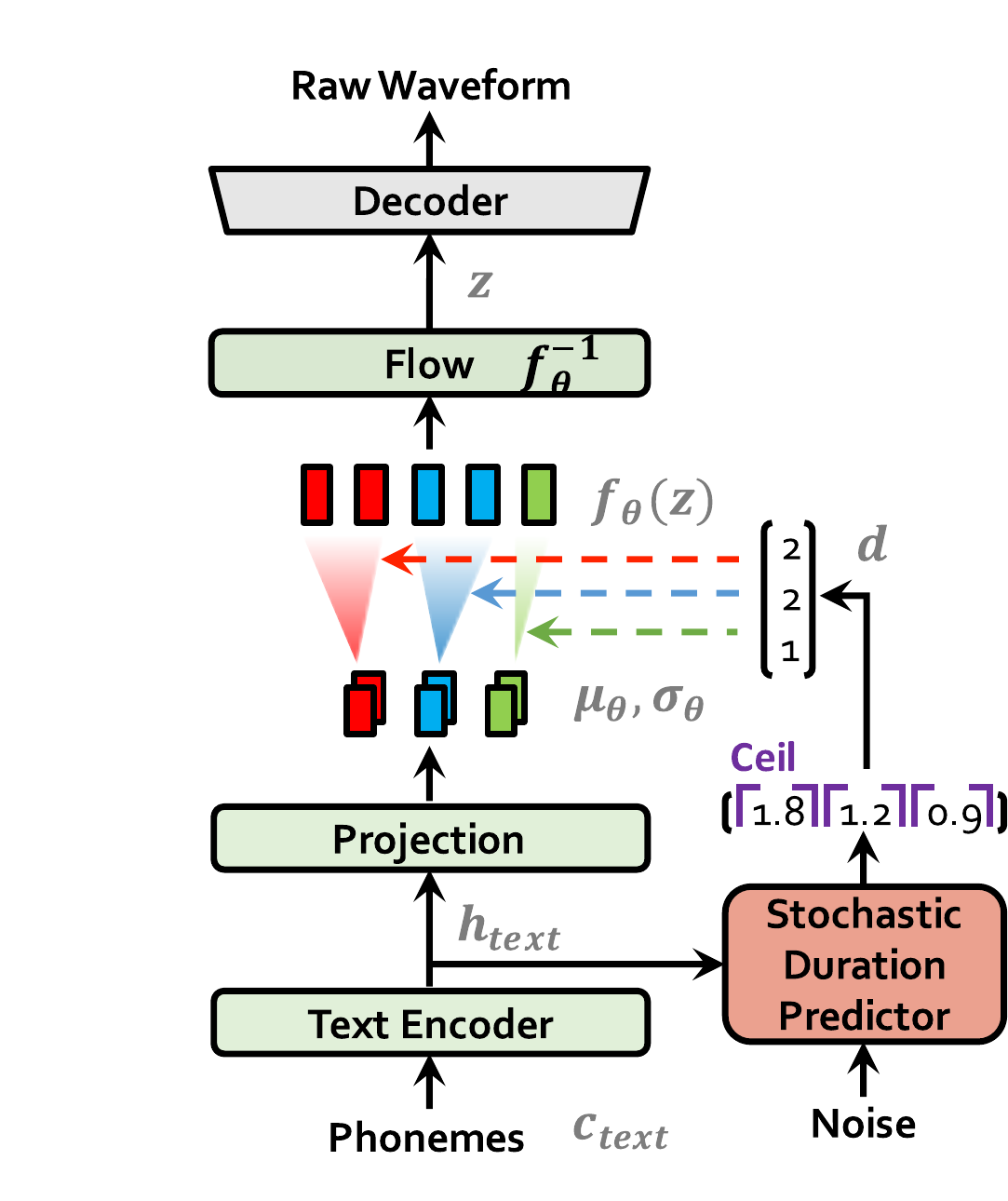}
            \vskip 0.1in
            \caption{Inference procedure}
            \label{fig:sampling}
        \end{subfigure}\hfill%
    
    \end{center}
    \vskip -0.05in
    \caption{System diagram depicting (a) training procedure and (b) inference procedure. The proposed model can be viewed as a conditional VAE; a posterior encoder, decoder, and conditional prior (green blocks: a normalizing flow, linear projection layer, and text encoder) with a flow-based stochastic duration predictor.}
\end{figure*}

In this section, we explain our proposed method and the architecture of it.
The proposed method is mostly described in the first three subsections: a conditional VAE formulation; alignment estimation derived from variational inference; adversarial training for improving synthesis quality. The overall architecture is described at the end of this section. Figures~\ref{fig:training} and \ref{fig:sampling} show the training and inference procedures of our method, respectively. From now on, we will refer to our method as Variational Inference with adversarial learning for end-to-end Text-to-Speech (VITS).

\subsection{Variational Inference}
\subsubsection{Overview}

VITS can be expressed as a conditional VAE with the objective of maximizing the variational lower bound, also called the evidence lower bound (ELBO), of the intractable marginal log-likelihood of data $\log{p_{\theta}(x|c)}$:
\begin{multline}
\label{eqn/cvae}
    \log{p_{\theta}(x|c)} \geq  
    \mathbb{E}_{q_{\phi}(z|x)}\Big[\log{p_{\theta}(x|z)} - \log{\frac{q_{\phi}(z|x)}{p_{\theta}(z|c)}} \Big]
\end{multline}
where $p_\theta(z|c)$ denotes a prior distribution of the latent variables $z$ given condition $c$, $p_\theta(x|z)$ is the likelihood function of a data point $x$, and $q_\phi(z|x)$ is an approximate posterior distribution. The training loss is then the negative ELBO, which can be viewed as the sum of reconstruction loss~$-\log{p_{\theta}(x|z)}$ and KL divergence~$\log{q_{\phi}(z|x)}-\log{p_{\theta}(z|c)}$, where $z \sim q_{\phi}(z|x)$.

\subsubsection{Reconstruction loss}

As a target data point in the reconstruction loss, we use a mel-spectrogram instead of a raw waveform, denoted by \(x_{mel}\). We upsample the latent variables \(z\) to the waveform domain \(\hat{y}\) through a decoder and transform \(\hat{y}\) to the mel-spectrogram domain \(\hat{x}_{mel}\). Then the \(L_1\) loss between the predicted and target mel-spectrogram is used as the reconstruction loss:
\begin{equation}
\label{eqn/recon}
    L_{recon} = \lVert x_{mel} - \hat{x}_{mel}\rVert_1
\end{equation}
This can be viewed as maximum likelihood estimation assuming a Laplace distribution for the data distribution and ignoring constant terms. We define the reconstruction loss in the mel-spectrogram domain to improve the perceptual quality by using a mel-scale that approximates the response of the human auditory system. Note that the mel-spectrogram estimation from a raw waveform does not require trainable parameters as it only uses STFT and linear projection onto the mel-scale. Furthermore, the estimation is only employed during training, not inference. In practice, we do not upsample the whole latent variables \(z\) but use partial sequences as an input for the decoder, which is the windowed generator training used for efficient end-to-end training~\cite{ren2021fastspeech,donahue2021endtoend}.

\subsubsection{KL-divergence}
The input condition of the prior encoder \(c\) is composed of phonemes \(c_{text}\) extracted from text and an alignment \(A\) between phonemes and latent variables. The alignment is a hard monotonic attention matrix with $|c_{text}| \times |z|$ dimensions representing how long each input phoneme expands to be time-aligned with the target speech. Because there are no ground truth labels for the alignment, we must estimate the alignment at each training iteration, which we will discuss in Section~\ref{sec/mas}. In our problem setting, we aim to provide more high-resolution information for the posterior encoder. We, therefore, use the linear-scale spectrogram of target speech \(x_{lin}\) as input rather than the mel-spectrogram. Note that the modified input does not violate the properties of variational inference. The KL divergence is then:
\begin{align}
\label{eqn/loss_kld}
    L_{kl} = \log{q_{\phi}(z|x_{lin})} - \log{p_{\theta}(z|c_{text},A)}, \\
    z \sim q_{\phi}(z|x_{lin}) = N(z;\mu_{\phi}(x_{lin}),\sigma_{\phi}(x_{lin})) \notag
\end{align}
The factorized normal distribution is used to parameterize our prior and posterior encoders. We found that increasing the expressiveness of the prior distribution is important for generating realistic samples. We, therefore, apply a normalizing flow $f_{\theta}$~\cite{rezende2015variational}, which allows an invertible transformation of a simple distribution into a more complex distribution following the rule of change-of-variables, on top of the factorized normal prior distribution:
\begin{align}
\label{eqn/prior}
    p_\theta(z|c) & = N(f_{\theta}(z);\mu_{\theta}(c),\sigma_{\theta}(c))\Big|\det\frac{\partial f_{\theta}(z)}{\partial z}\Big|, \\
    c & =[c_{text}, A] \notag
\end{align}

\subsection{Alignment Estimation}

\subsubsection{Monotonic alignment search}
\label{sec/mas}

To estimate an alignment $A$ between input text and target speech, we adopt Monotonic Alignment Search (MAS)~\cite{kim2020glow}, a method to search an alignment that maximizes the likelihood of data parameterized by a normalizing flow $f$:
\begin{align}
\label{eqn/mas1}
A & = \argmax\limits_{\hat{A}}\ {\log{p(x|c_{text},\hat{A})}}\notag \\
& = \argmax\limits_{\hat{A}}\ {\log{N(f(x);\mu(c_{text},\hat{A}), \sigma(c_{text},\hat{A}))}}
\end{align}
where the candidate alignments are restricted to be monotonic and non-skipping following the fact that humans read text in order without skipping any words. 
To find the optimum alignment, \citet{kim2020glow} use dynamic programming. Applying MAS directly in our setting is difficult because our objective is the ELBO, not the exact log-likelihood. We, therefore, redefine MAS to find an alignment that maximizes the ELBO, which reduces to finding an alignment that maximizes the log-likelihood of the latent variables $z$:
\begin{flalign}
\label{eqn/mas2}
 &\argmax\limits_{\hat{A}}\ {\log{p_{\theta}(x_{mel}|z)} - \log{\frac{q_{\phi}(z|x_{lin})}{p_{\theta}(z|c_{text},\hat{A})}}}\notag &&\\
 &= \argmax\limits_{\hat{A}}\ {\log{p_{\theta}(z|c_{text},\hat{A})}}\notag &&\\
 &\;\;\;\;\;\;\;\;\;\;\;\;= \log{N(f_{\theta}(z);\mu_{\theta}(c_{text},\hat{A}),\sigma_{\theta}(c_{text},\hat{A}))} &&
\end{flalign}
Due to the resemblance of Equation~\ref{eqn/mas1} to Equation~\ref{eqn/mas2}, we can use the original MAS implementation without modification. Appendix~\hyperref[app:mas]{A} includes pseudocode for MAS.

\subsubsection{Duration prediction from text}
\label{sec/dp}
We can calculate the duration of each input token $d_{i}$ by summing all the columns in each row of the estimated alignment $\sum_{j} A_{i,j}$. The duration could be used to train a deterministic duration predictor, as proposed in previous work~\cite{kim2020glow}, but it cannot express the way a person utters at different speaking rates each time. To generate human-like rhythms of speech, we design a stochastic duration predictor so that its samples follow the duration distribution of given phonemes. The stochastic duration predictor is a flow-based generative model that is typically trained via maximum likelihood estimation. The direct application of maximum likelihood estimation, however, is difficult because the duration of each input phoneme is 1) a discrete integer, which needs to be dequantized for using continuous normalizing flows, and 2) a scalar, which prevents high-dimensional transformation due to invertibility. We apply variational dequantization~\cite{ho2019flow++} and variational data augmentation~\cite{chen2020vflow} to solve these problems. To be specific, we introduce two random variables $u$ and $\nu$, which have the same time resolution and dimension as that of the duration sequence $d$, for variational dequatization and variational data augmentation, respectively. We restrict the support of $u$ to be $[0,1)$ so that the difference $d - u$ becomes a sequence of positive real numbers, and we concatenate $\nu$ and $d$ channel-wise to make a higher dimensional latent representation. We sample the two variables through an approximate posterior distribution $q_{\phi}(u,\nu|d,c_{text})$.
The resulting objective is a variational lower bound of the log-likelihood of the phoneme duration:
\begin{multline}
\label{eqn/sdp}
\log{p_{\theta}(d|c_{text})} \geq \\ \mathbb{E}_{q_{\phi}(u,\nu|d,c_{text})}\Big[\log{\frac{p_{\theta}(d-u,\nu|c_{text})}{q_{\phi}(u,\nu|d,c_{text})}}\Big]
\end{multline}
The training loss $L_{dur}$ is then the negative variational lower bound. We apply the stop gradient operator~\cite{van2017neural}, which prevents back-propagating the gradient of inputs, to the input conditions so that the training of the duration predictor does not affect that of other modules.

The sampling procedure is relatively simple; the phoneme duration is sampled from random noise through the inverse transformation of the stochastic duration predictor, and then it is converted to integers.

\subsection{Adversarial Training}
To adopt adversarial training in our learning system, we add a discriminator $D$ that distinguishes between the output generated by the decoder $G$ and the ground truth waveform $y$. In this work, we use two types of loss successfully applied in speech synthesis; the least-squares loss function~\cite{mao2017least} for adversarial training, and the additional feature-matching loss~\cite{larsen2016autoencoding} for training the generator:
\begin{align}
    L_{adv}(D) &= \mathbb{E}_{(y,z)}\Big[(D(y)-1)^2+(D(G(z)))^2\Big], \\
    L_{adv}(G) &= \mathbb{E}_z\Big[(D(G(z))-1)^2\Big], \\
    L_{fm}(G) &= \mathbb{E}_{(y,z)}\Big[\sum_{l=1}^{T}\frac{1}{N_l}\lVert D^{l}(y) - D^{l}(G(z)) \rVert_1\Big] 
\end{align}
where $T$ denotes the total number of layers in the discriminator and $D^{l}$ outputs the feature map of the $l$-th layer of the discriminator with $N_{l}$ number of features.
Notably, the feature matching loss can be seen as reconstruction loss that is measured in the hidden layers of the discriminator suggested as an alternative to the element-wise reconstruction loss of VAEs~\cite{larsen2016autoencoding}.

\subsection{Final Loss}

With the combination of VAE and GAN training, the total loss for training our conditional VAE can be expressed as follows:
\begin{equation}
  \label{eqn/total_loss}
    L_{vae} =  L_{recon} + L_{kl} + L_{dur} + L_{adv}(G) + L_{fm}(G)
\end{equation}

\subsection{Model Architecture}
\label{model_architecture}
The overall architecture of the proposed model consists of a posterior encoder, prior encoder, decoder, discriminator, and stochastic duration predictor. The posterior encoder and discriminator are only used for training, not for inference. Architectural details are available in Appendix~\hyperref[app:arch]{B}.

\subsubsection{Posterior Encoder} 
For the posterior encoder, we use the non-causal WaveNet residual blocks used in WaveGlow~\cite{prenger2019waveglow} and Glow-TTS~\cite{kim2020glow}. A WaveNet residual block consists of layers of dilated convolutions with a gated activation unit and skip connection. The linear projection layer above the blocks produces the mean and variance of the normal posterior distribution. For the multi-speaker case, we use global conditioning~\cite{oord2016wavenet} in residual blocks to add speaker embedding.

\subsubsection{Prior Encoder}
The prior encoder consists of a text encoder that processes the input phonemes \(c_{text}\) and a normalizing flow $f_{\theta}$ that improves the flexibility of the prior distribution. The text encoder is a transformer encoder~\cite{vaswani2017attention} that uses relative positional representation~\cite{shaw2018self} instead of absolute positional encoding. We can obtain the hidden representation \(h_{text}\) from \(c_{text}\) through the text encoder and a linear projection layer above the text encoder that produces the mean and variance used for constructing the prior distribution. The normalizing flow is a stack of affine coupling layers~\cite{dinh2016density} consisting of a stack of WaveNet residual blocks. For simplicity, we design the normalizing flow to be a volume-preserving transformation with the Jacobian determinant of one. For the multi-speaker setting, we add speaker embedding to the residual blocks in the normalizing flow through global conditioning.

\subsubsection{Decoder}
The decoder is essentially the HiFi-GAN V1 generator~\cite{kong2020hifi}. It is composed of a stack of transposed convolutions, each of which is followed by a multi-receptive field fusion module (MRF). The output of the MRF is the sum of the output of residual blocks that have different receptive field sizes. For the multi-speaker setting, we add a linear layer that transforms speaker embedding and add it to the input latent variables $z$.

\subsubsection{Discriminator}
We follow the discriminator architecture of the multi-period discriminator proposed in HiFi-GAN~\cite{kong2020hifi}. The multi-period discriminator is a mixture of Markovian window-based sub-discriminators~\cite{kumar2019melgan}, each of which operates on different periodic patterns of input waveforms.

\subsubsection{Stochastic Duration Predictor}
The stochastic duration predictor estimates the distribution of phoneme duration from a conditional input \(h_{text}\). For the efficient parameterization of the stochastic duration predictor, we stack residual blocks with dilated and depth-separable convolutional layers. We also apply neural spline flows~\cite{durkan2019neural}, which take the form of invertible nonlinear transformations by using monotonic rational-quadratic splines, to coupling layers. Neural spline flows improve transformation expressiveness with a similar number of parameters compared to commonly used affine coupling layers. For the multi-speaker setting, we add a linear layer that transforms speaker embedding and add it to the input $h_{text}$.

%% file: sections/experiments.tex
\section{Experiments}
\subsection{Datasets}
\label{sec_exp_datasets}
We conducted experiments on two different datasets. We used the LJ Speech dataset~\citep{ljspeech17} for comparison with other publicly available models and the VCTK dataset~\citep{veaux2017cstr} to verify whether our model can learn and express diverse speech characteristics. The LJ Speech dataset consists of 13,100 short audio clips of a single speaker with a total length of approximately 24 hours. The audio format is 16-bit PCM with a sample rate of 22 kHz, and we used it without any manipulation. We randomly split the dataset into a training set (12,500 samples), validation set (100 samples), and test set (500 samples). The VCTK dataset consists of approximately 44,000 short audio clips uttered by 109 native English speakers with various accents. The total length of the audio clips is approximately 44 hours. The audio format is 16-bit PCM with a sample rate of 44 kHz. We reduced the sample rate to 22 kHz. We randomly split the dataset into a training set (43,470 samples), validation set (100 samples), and test set (500 samples).

\subsection{Preprocessing}
We use linear spectrograms which can be obtained from raw waveforms through the Short-time Fourier transform (STFT), as input of the posterior encoder. The FFT size, window size and hop size of the transform are set to 1024, 1024 and 256, respectively. We use 80 bands mel-scale spectrograms for reconstruction loss, which is obtained by applying a mel-filterbank to linear spectrograms.

We use International Phonetic Alphabet (IPA) sequences as input to the prior encoder. We convert text sequences to IPA phoneme sequences using open-source software~\citep{phonemizer20}, and the converted sequences are interspersed with a blank token following the implementation of Glow-TTS.

\subsection{Training}
The networks are trained using the AdamW optimizer~\citep{loshchilov2018decoupled} with $\beta_1 = 0.8$, $\beta_2 = 0.99$ and weight decay $\lambda = 0.01$. The learning rate decay is scheduled by a $0.999^{1/8}$ factor in every epoch with an initial learning rate of $2\times10^{-4}$.
Following previous work~\citep{ren2021fastspeech, donahue2021endtoend}, we adopt the windowed generator training, a method of generating only a part of raw waveforms to reduce the training time and memory usage during training. We randomly extract segments of latent representations with a window size of 32 to feed to the decoder instead of feeding entire latent representations and also extract the corresponding audio segments from the ground truth raw waveforms as training targets.
We use mixed precision training on 4 NVIDIA V100 GPUs. The batch size is set to 64 per GPU and the model is trained up to 800k steps.

\subsection{Experimental Setup for Comparison}
We compared our model with the best publicly available models. We used Tacotron 2, an autoregressive model, and Glow-TTS, a flow-based non-autoregressive model, as first stage models and HiFi-GAN as a second stage model. We used their public implementations and pre-trained weights.\footnote{The implementations are as follows:\\
Tacotron 2 : \href{https://github.com/NVIDIA/tacotron2}{https://github.com/NVIDIA/tacotron2}\\
Glow-TTS : \href{https://github.com/jaywalnut310/glow-tts}{https://github.com/jaywalnut310/glow-tts}\\
HiFi-GAN : \href{https://github.com/jik876/hifi-gan}{https://github.com/jik876/hifi-gan}} 
Since a two-stage TTS system can theoretically achieve higher synthesis quality through sequential training, we included the fine-tuned HiFi-GAN up to 100k steps with the predicted outputs from the first stage models. We empirically found that fine-tuning HiFi-GAN with the generated mel-spectrograms from Tacotron 2 under teacher-forcing mode, led to better quality for both Tacotron 2 and Glow-TTS than fine-tuning with the generated mel-spectrograms from Glow-TTS, so we appended the better fine-tuned HiFi-GAN to both Tacotron 2 and Glow-TTS.

As each model has a degree of randomness during sampling, we fixed hyper-parameters that controls the randomness of each model throughout our experiments. The probability of dropout in the pre-net of Tactron 2 was set to 0.5. For Glow-TTS, the standard deviation of the prior distribution was set to 0.333. For VITS, the standard deviation of input noise of the stochastic duration predictor was set to 0.8 and we multiplied a scale factor of 0.667 to the standard deviation of the prior distribution.

%% file: sections/results.tex
\section{Results}
\subsection{Speech Synthesis Quality}
\label{sec_res_quality}
We conducted crowd-sourced MOS tests to evaluate the quality. Raters listened to randomly selected audio samples, and rated their naturalness on a 5 point scale from 1 to 5. Raters were allowed to evaluate each audio sample once, and we normalized all the audio clips to avoid the effect of amplitude differences on the score. All of the quality assessments in this work were conducted in this manner.

The evaluation results are shown in Table~\ref{tab:mos-comparison}.
VITS outperforms other TTS systems and achieves a similar MOS to that of ground truth. The VITS (\textbf{DDP}), which employs the same deterministic duration predictor architecture used in Glow-TTS rather than the stochastic duration predictor, scores the second-highest among TTS systems in the MOS evaluation. These results imply that 1) the stochastic duration predictor generates more realistic phoneme duration than the deterministic duration predictor and 2) our end-to-end training method is an effective way to make better samples than other TTS models even if maintaining the similar duration predictor architecture.

\begin{table}[h]
  \caption{Comparison of evaluated MOS with 95\% confidence intervals on the LJ Speech dataset.}
  \label{tab:mos-comparison}
  \vskip 0.15in
  \begin{center}
  \begin{tabular}{lc}
    \toprule
    Model   & 
    MOS (CI) \\
    \midrule
    Ground Truth    & 4.46 ($\pm$0.06) \\
    Tacotron 2 + HiFi-GAN & 3.77 ($\pm$0.08) \\
    Tacotron 2 + HiFi-GAN (Fine-tuned) & 4.25 ($\pm$0.07) \\
    Glow-TTS + HiFi-GAN & 4.14 ($\pm$0.07) \\
    Glow-TTS + HiFi-GAN (Fine-tuned) & 4.32 ($\pm$0.07) \\
    VITS (DDP) & 4.39 ($\pm$0.06) \\
    VITS & \textbf{4.43 ($\pm$0.06)} \\
    \bottomrule
  \end{tabular}
  \end{center}
  \vskip -0.10in
\end{table}

We conducted an ablation study to demonstrate the effectiveness of our methods, including the normalized flow in the prior encoder and linear-scale spectrogram posterior input. All models in the ablation study were trained up to 300k steps. The results are shown in Table~\ref{tab:ablation-studies}. Removing the normalizing flow in the prior encoder results in a 1.52 MOS decrease from the baseline, demonstrating that the prior distribution's flexibility significantly influences the synthesis quality. Replacing the linear-scale spectrogram for posterior input with the mel-spectrogram results in a  quality degradation (-0.19 MOS), indicating that the high-resolution information is effective for VITS in improving the synthesis quality.

\begin{table}[ht]
  \caption{MOS comparison in the ablation studies. }
  \label{tab:ablation-studies}
  \vskip 0.15in
  \begin{center}
  \begin{tabular}[htbp]{lc}
    \toprule
    Model   & 
    MOS (CI) \\
    \midrule
    Ground Truth    & 4.50 ($\pm$0.06) \\
    Baseline    & 4.50 ($\pm$0.06) \\
    without Normalizing Flow & 2.98 ($\pm$0.08) \\
    with Mel-spectrogram & 4.31 ($\pm$0.08) \\
    \bottomrule
  \end{tabular}
  \end{center}
  \vskip -0.1in
\end{table}

\subsection{Generalization to Multi-Speaker Text-to-Speech}

To verify that our model can learn and express diverse speech characteristics, we compared our model to Tacotron 2, Glow-TTS and HiFi-GAN, which showed the ability to extend to multi-speaker speech synthesis~\citep{jia2018transfer, kim2020glow, kong2020hifi}. We trained the models on the VCTK dataset. We added speaker embedding to our model as described in Section~\ref{model_architecture}. For Tacotron 2, we broadcasted speaker embedding and concatenated it with the encoder output, and for Glow-TTS, we applied the global conditioning following the previous work. The evaluation method is the same as that described in Section~\ref{sec_res_quality}. As shown in Table~\ref{tab:mos-ms-comparison}, our model achieves a higher MOS than the other models. This demonstrates that our model learns and expresses various speech characteristics in an end-to-end manner.
\begin{table}[ht]
  \caption{Comparison of evaluated MOS with 95\% confidence intervals on the VCTK dataset.}
  \label{tab:mos-ms-comparison}
  \vskip 0.15in
  \begin{center}
  \begin{tabular}{lc}
    \toprule
    Model   & 
    MOS (CI) \\
    \midrule
    Ground Truth    & 4.38 ($\pm$0.07) \\
    Tacotron 2 + HiFi-GAN & 3.14 ($\pm$0.09) \\
    Tacotron 2 + HiFi-GAN (Fine-tuned) & 3.19 ($\pm$0.09) \\
    Glow-TTS + HiFi-GAN & 3.76 ($\pm$0.07) \\
    Glow-TTS + HiFi-GAN (Fine-tuned) & 3.82 ($\pm$0.07) \\
    VITS & \textbf{4.38 ($\pm$0.06)} \\
    \bottomrule
  \end{tabular}
  \end{center}
  \vskip -0.1in
\end{table}

\subsection{Speech Variation}
\begin{figure}[ht]
    \begin{center}
        \begin{subfigure}{.49\textwidth}
            \centering
            \includegraphics[width=1.\linewidth]{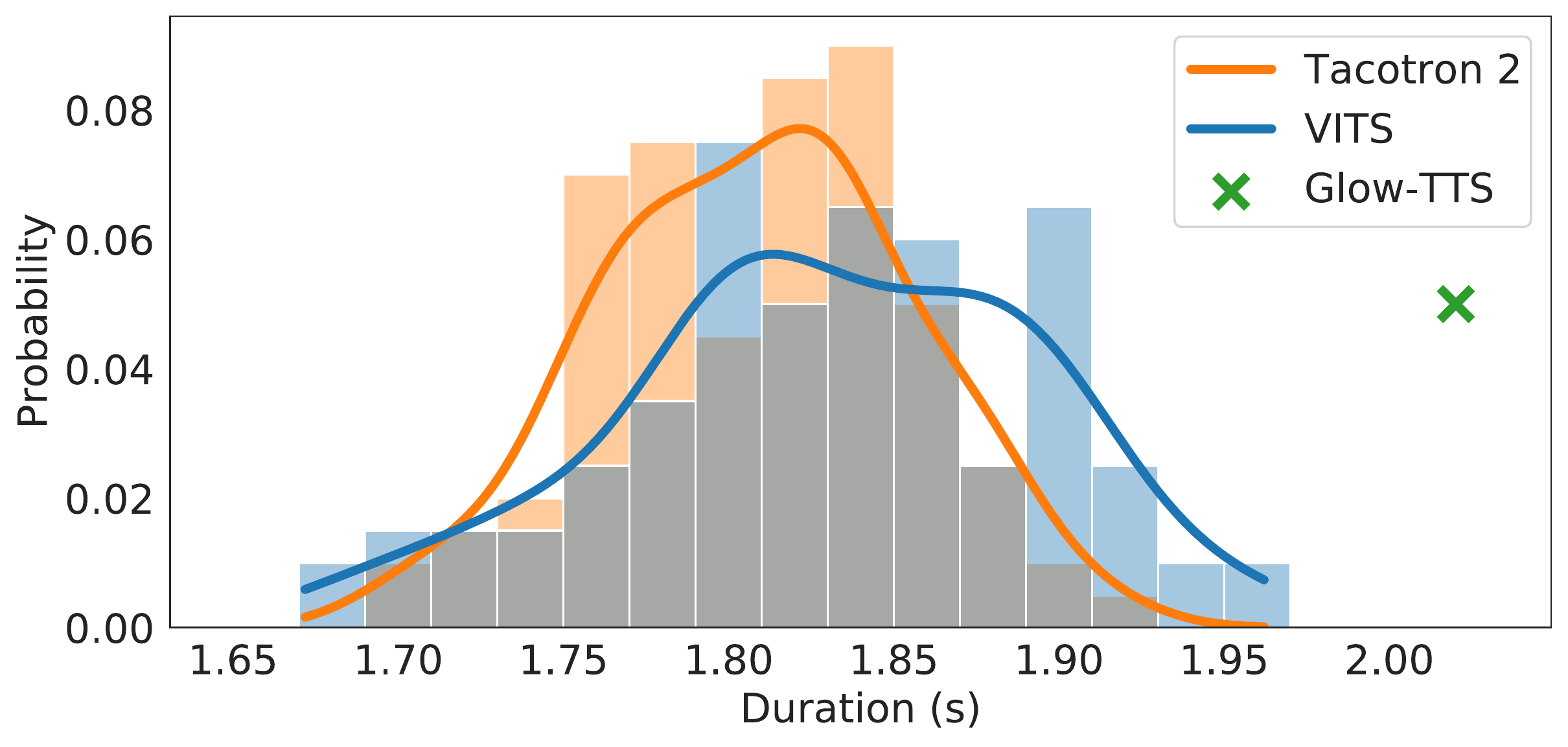}
            \caption{Comparison of sample duration. Glow-TTS only provides a single value due to the deterministic duration predictor.}
            \vskip 0.1in
            \label{fig:dur_ss}
        \end{subfigure}\hfill%
        \begin{subfigure}{.49\textwidth}
            \centering
            \includegraphics[width=1.\linewidth]{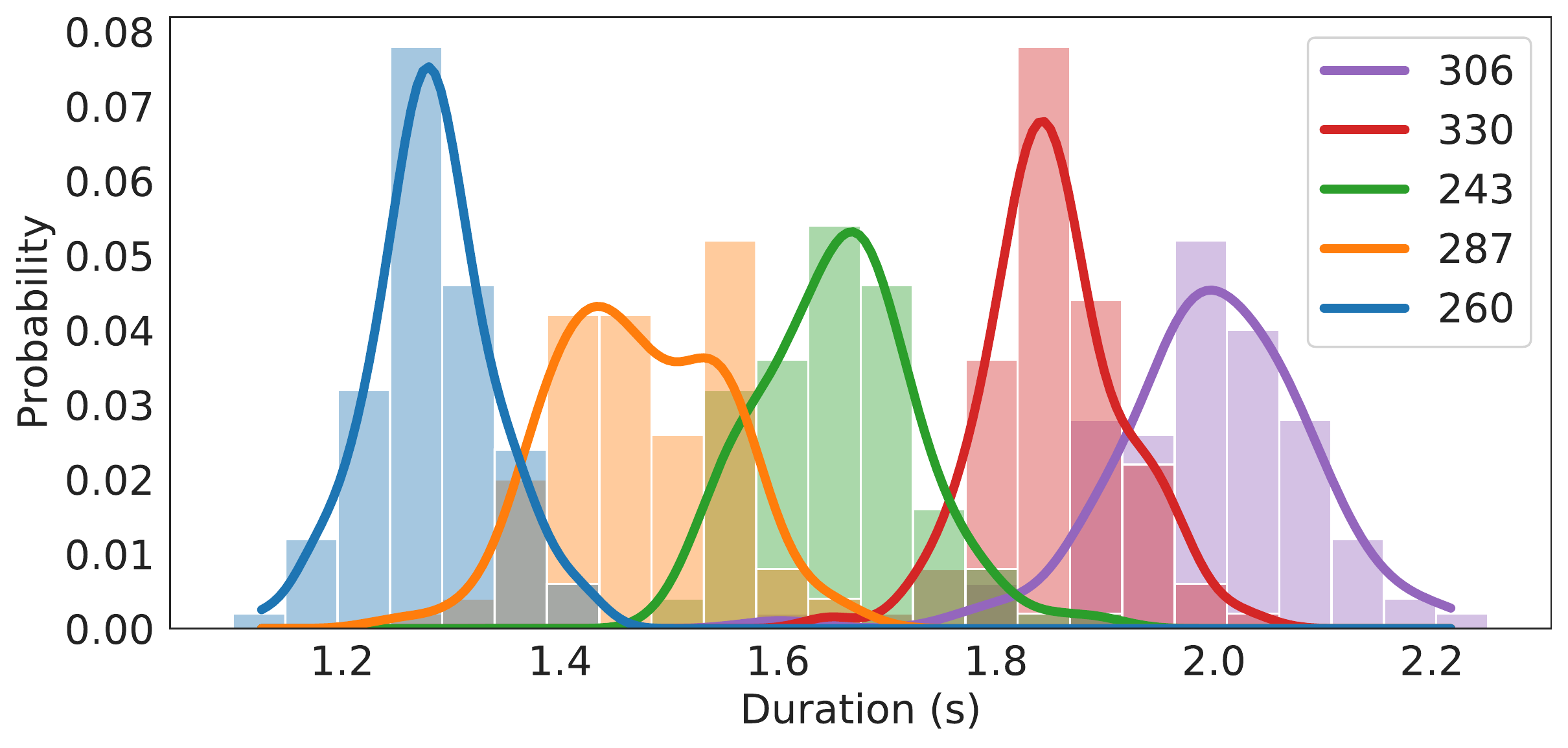}
            \caption{Comparison of sample duration in different speakers.}
            \label{fig:dur_ms}
        \end{subfigure}\hfill%
    \end{center}
    \vskip -0.05in
    \caption{Sample duration in seconds on (a) the LJ Speech dataset and (b) the VCTK dataset.}
\end{figure}

\begin{figure}[ht]
    \vskip 0.2in
    \begin{center}
        \begin{subfigure}{.49\textwidth}
            \centering
            \includegraphics[width=1.\linewidth]{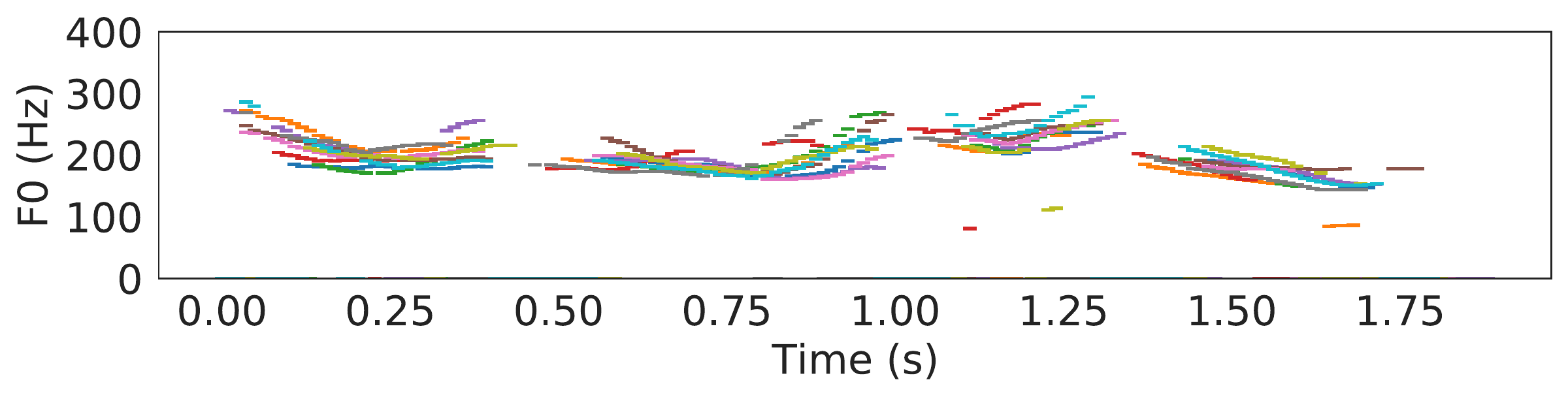}
            \caption{VITS}
            \vskip 0.05in
            \label{fig:f0_etts}
        \end{subfigure}\hfill%
        \begin{subfigure}{.49\textwidth}
            \centering
            \includegraphics[width=1.\linewidth]{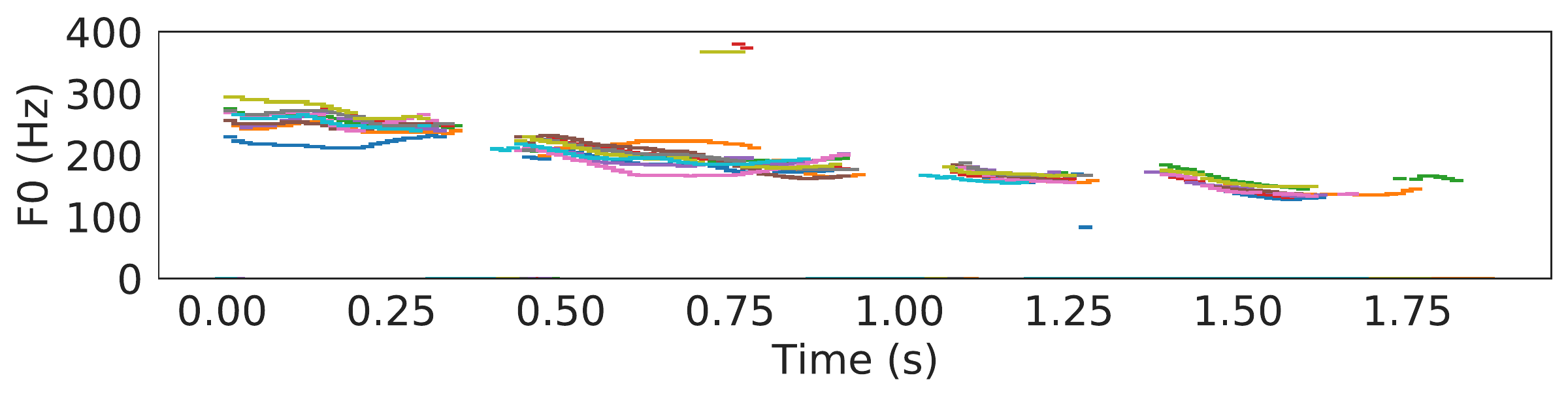}
            \caption{Tacotron 2}
            \vskip 0.05in
            \label{fig:f0_taco}
        \end{subfigure}\hfill%
        \begin{subfigure}{.49\textwidth}
            \centering
            \includegraphics[width=1.\linewidth]{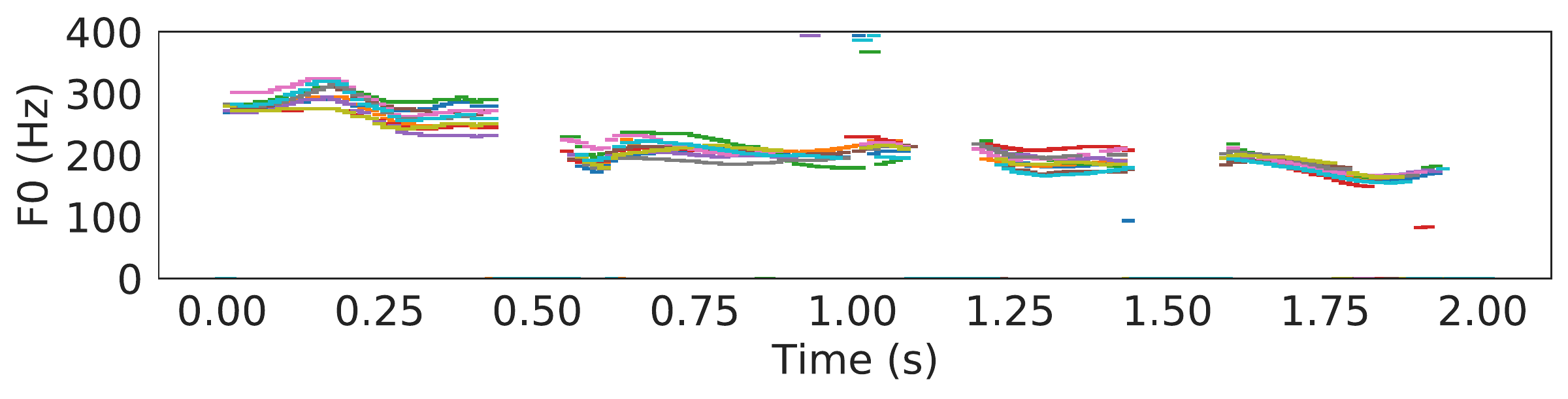}
            \caption{Glow-TTS}
            \vskip 0.05in
            \label{fig:f0_gtts}
        \end{subfigure}\hfill%
        \begin{subfigure}{.49\textwidth}
            \centering
            \includegraphics[width=1.\linewidth]{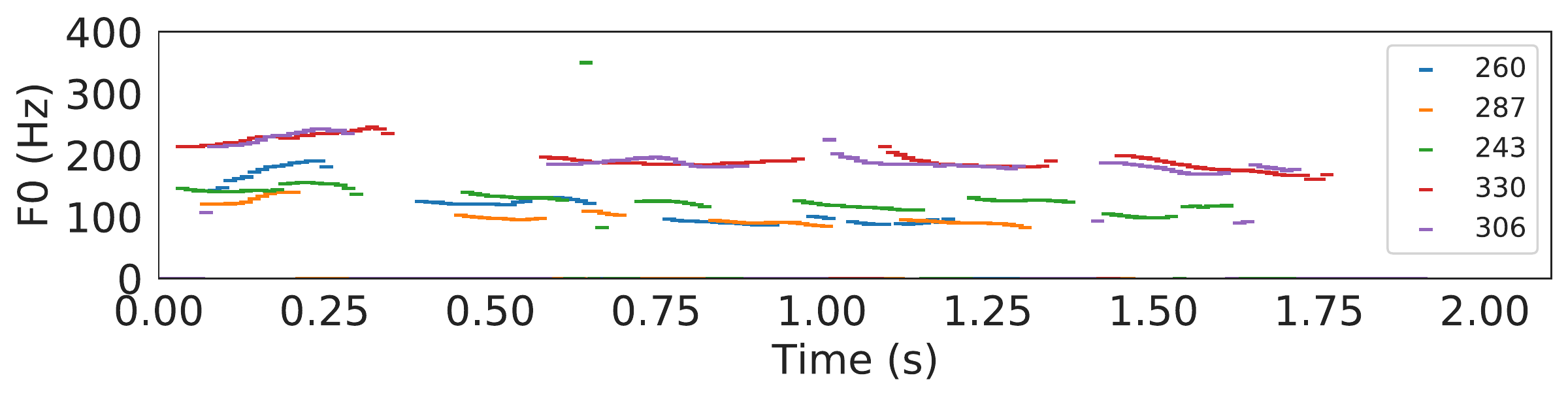}
            \caption{VITS (multi-speaker)}
            \vskip 0.05in
            \label{fig:f0_etts_ms}
        \end{subfigure}\hfill%
    \end{center}
    \vskip -0.05in
    \caption{Pitch tracks for the utterance ``How much variation is there?”. Samples are generated from (a) VITS, (b) Tacotron 2, and (c) Glow-TTS in the single speaker setting and from (d) VITS in the multi-speaker setting.}
    \label{fig:f0}
\end{figure}

We verified how many different lengths of speech the stochastic duration predictor produces, and how many different speech characteristics the synthesized samples have. Similar to \citet{valle2021flowtron}, all samples here were generated from a sentence ``\textit{How much variation is there?}". Figure~\ref{fig:dur_ss} shows histograms of the lengths of 100 generated utterances from each model. While Glow-TTS generates only fixed-length utterances due to the deterministic duration predictor, samples from our model follow a similar length distribution to that of Tacotron 2. Figure~\ref{fig:dur_ms} shows the lengths of 100 utterances generated with each of five speaker identities from our model in the multi-speaker setting, implying that the model learns the speaker-dependent phoneme duration. F0 contours of 10 utterances extracted with the YIN algorithm~\cite{de2002yin} in Figure~\ref{fig:f0} shows that our model generates speech with diverse pitches and rhythms, and five samples generated with each of different speaker identities in Figure~\ref{fig:f0_etts_ms} demonstrates our model expresses very different lengths and pitches of speech for each speaker identity. Note that Glow-TTS could increase the diversity of pitch by increasing the standard deviation of the prior distribution, but on the contrary, it could lower the synthesis quality.

\subsection{Synthesis Speed}
We compared the synthesis speed of our model with a parallel two-stage TTS system, Glow-TTS and HiFi-GAN. We measured the synchronized elapsed time over the entire process to generate raw waveforms from phoneme sequences with 100 sentences randomly selected from the test set of the LJ Speech dataset. We used a single NVIDIA V100 GPU with a batch size of 1. The results are shown in Table~\ref{tab:synthesis_speed}. Since our model does not require modules for generating predefined intermediate representations, its sampling efficiency and speed are greatly improved.
\begin{table}[ht]
  \caption{Comparison of the synthesis speed. Speed of $n$~kHz means that the model can generate $n\times1000$ raw audio samples per second. Real-time means the synthesis speed over real-time.}
  \label{tab:synthesis_speed}
  \vskip 0.15in
  \begin{center}
  \begin{tabular}{lcc}
    \toprule
    Model   & 
    Speed (kHz) & Real-time \\
    \midrule
    Glow-TTS + HiFi-GAN & 606.05 & $\times$27.48 \\
    VITS & 1480.15 & $\times$67.12 \\
    VITS (DDP) & 2005.03 & $\times$90.93 \\
    \bottomrule
  \end{tabular}
  \end{center}
  \vskip -0.1in
\end{table}

%% file: sections/relatedwork.tex
\section{Related Work}
\subsection{End-to-End Text-to-Speech}
Currently, neural TTS models with a two-stage pipeline can synthesize human-like speech~\cite{oord2016wavenet, ping2018deep, shen2018natural}. However, they typically require vocoders trained or fine-tuned with first stage model output, which causes training and deployment inefficiency. They are also unable to reap the potential benefits of an end-to-end approach that can use learned hidden representations rather than predefined intermediate features.

Recently, single-stage end-to-end TTS models have been proposed to tackle the more challenging task of generating raw waveforms, which contain richer information (e.g., high-frequency response and phase) than mel-spectrograms, directly from text. FastSpeech 2s~\cite{ren2021fastspeech} is an extension of FastSpeech 2 that enables end-to-end parallel generation by adopting adversarial training and an auxiliary mel-spectrogram decoder that helps learn text representations. However, to resolve the one-to-many problem, FastSpeech 2s must extract phoneme duration, pitch, and energy from speech used as input conditions in training. EATS~\cite{donahue2021endtoend} employs adversarial training as well and a differentiable alignment scheme. To resolve the length mismatch problem between generated and target speech, EATS adopts soft dynamic time warping loss that is calculated by dynamic programming. Wave Tacotron~\cite{weiss2020wave} combines normalizing flows with Tacotron 2 for an end-to-end structure but remains autoregressive. The audio quality of all the aforementioned end-to-end TTS models is less than that of two-stage models.

Unlike the aforementioned end-to-end models, by utilizing a conditional VAE, our model 1) learns to synthesize raw waveforms directly from text without requiring additional input conditions, 2) uses a dynamic programming method, MAS, to search the optimal alignment rather than to calculate loss, 3) generates samples in parallel, and 4) outperforms the best publicly available two-stage models.

\subsection{Variational Autoencoders}
VAEs~\cite{kingma2013auto} are one of the most widely used likelihood-based deep generative models. We adopt a conditional VAE to a TTS system. A conditional VAE is a conditional generative model where the observed conditions modulate the prior distribution of latent variables used to generate outputs. In speech synthesis, \citet{hsu2018hierarchical} and \citet{zhang2019learning} combine Tacotron 2 and VAEs to learn speech style and prosody. BVAE-TTS~\cite{lee2021bidirectional} generates mel-spectrograms in parallel based on a bidirectional VAE~\cite{kingma2016improved}. Unlike the previous works that applied VAEs to first stage models, we adopt a VAE to a parallel end-to-end TTS system.

\citet{rezende2015variational}, \citet{chen2016variational} and \citet{ziegler2019latent} improve VAE performance by enhancing the expressive power of prior and posterior distribution with normalizing flows. To improve the representation power of the prior distribution, we add normalizing flows to our conditional prior network, leading to the generation of more realistic samples.

Similar to our work, \citet{ma2019flowseq} proposed a conditional VAE with normalizing flows in a conditional prior network for non-autoregressive neural machine translation, FlowSeq. However, the fact that our model can explicitly align a latent sequence with the source sequence differs from FlowSeq, which needs to learn implicit alignment through attention mechanisms. Our model removes the burden of transforming the latent sequence into standard normal random variables by matching the latent sequence with the time-aligned source sequence via MAS, which allows for simpler architecture of normalizing flows. 

\subsection{Duration Prediction in Non-Autoregressive Text-to-Speech}
Autoregressive TTS models~\cite{taigman2018voiceloop, shen2018natural, valle2021flowtron} generate diverse speech with different rhythms through their autoregressive structure and several tricks including maintaining dropout probability during inference and priming~\cite{graves2013generating}. Parallel TTS models~\cite{ren2019fastspeech, peng2020non, kim2020glow, ren2021fastspeech, lee2021bidirectional}, on the other hand, have been relied on deterministic duration prediction. It is because parallel models have to predict target phoneme duration or the total length of target speech in one feed-forward path, which makes it hard to capture the correlated joint distribution of speech rhythms. In this work, we suggest a flow-based stochastic duration predictor that learns the joint distribution of the estimated phoneme duration, resulting in the generation of diverse speech rhythms in parallel.

%% file: sections/conclusion.tex
\section{Conclusion}
In this work, we proposed a parallel TTS system, VITS, that can learn and generate in an end-to-end manner. We further introduced the stochastic duration predictor to express diverse rhythms of speech. The resulting system synthesizes natural sounding speech waveforms directly from text, without having to go through predefined intermediate speech representations. Our experimental results show that our method outperforms two-stage TTS systems and achieves close to human quality. 
We hope the proposed method will be used in many speech synthesis tasks, where two-stage TTS systems have been used, to achieve performance improvement and enjoy the simplified training procedure. We also want to point out that even though our method integrates two separated generative pipelines in TTS systems, there remains a problem of text preprocessing. Investigating self-supervised learning of language representations could be a possible direction for removing the text preprocessing step. We will release our source-code and pre-trained models to facilitate research in plenty of future directions.

%% file: sections/acknowledgements.tex
\section*{Acknowledgements}
We would like to thank Sungwon Lyu, Bokyung Son, Sunghyo Chung, and Jonghoon Mo for helpful discussions and advice.

%% file: sections/appendix.tex
\newpage
\onecolumn
\icmltitle{Supplementary Material of \\Conditional Variational Autoencoder with Adversarial Learning for End-to-End Text-to-Speech}

\section*{A. Monotonic Alignment Search}
\label{app:mas}
We present pseudocode for MAS in Figure~\ref{fig:mas}. Although we search the alignment which maximizes the ELBO not the exact log-likelihood of data, we can use the MAS implementation of Glow-TTS as described in Section~\hyperref[sec/mas]{2.2.1}.

\begin{figure*}[h]
\centering
\scalebox{0.80}{
    \begin{minipage}{1.0\textwidth}
        \input{figures/mas_pseudocode.tex}
    \end{minipage}
}
\caption{Pseudocode for Monotonic Alignment Search.}
\label{fig:mas}
\end{figure*}

\section*{B. Model Configurations}
\label{app:arch}
In this section, we mainly describe the newly added parts of VITS as we followed configurations of Glow-TTS and HiFi-GAN for several parts of our model: we use the same transformer encoder and WaveNet residual blocks as those of Glow-TTS; our decoder and the multi-period discriminator is the same as the generator and multi-period discriminator of HiFi-GAN, respectively, except that we use different input dimension for the decoder and append a sub-discriminator.

\subsection*{B.1. Prior Encoder and Posterior Encoder}
The normalizing flow in the prior encoder is a stack of four affine coupling layers, each coupling layer consisting of four WaveNet residual blocks. As we restrict the affine coupling layers to be volume-preserving transformations, the coupling layers do not produce scale parameters.

The posterior encoder, consisting of 16 WaveNet residual blocks, takes linear-scale log magnitude spectrograms and produce latent variables with 192 channels.

\subsection*{B.2. Decoder and Discriminator} The input of our decoder is latent variables generated from the prior or posterior encoders, so the input channel size of the decoder is 192. For the last convolutional layer of the decoder, we remove a bias parameter, as it causes unstable gradient scales during mixed precision training. 

For the discriminator, HiFi-GAN uses the multi-period discriminator containing five sub-discriminators with periods $[2, 3, 5, 7, 11]$ and the multi-scale discriminator containing three sub-discriminators. To improve training efficiency, we leave only the first sub-discriminator of the multi-scale discriminator that operates on raw waveforms and discard two sub-discriminators operating on average-pooled waveforms. The resultant discriminator can be seen as the multi-period discriminator with periods $[1, 2, 3, 5, 7, 11]$.

\begin{figure*}[h]
    \vskip 0.2in
    \begin{center}
        \begin{subfigure}{.32\textwidth}
            \centering
            \includegraphics[width=1.\linewidth]{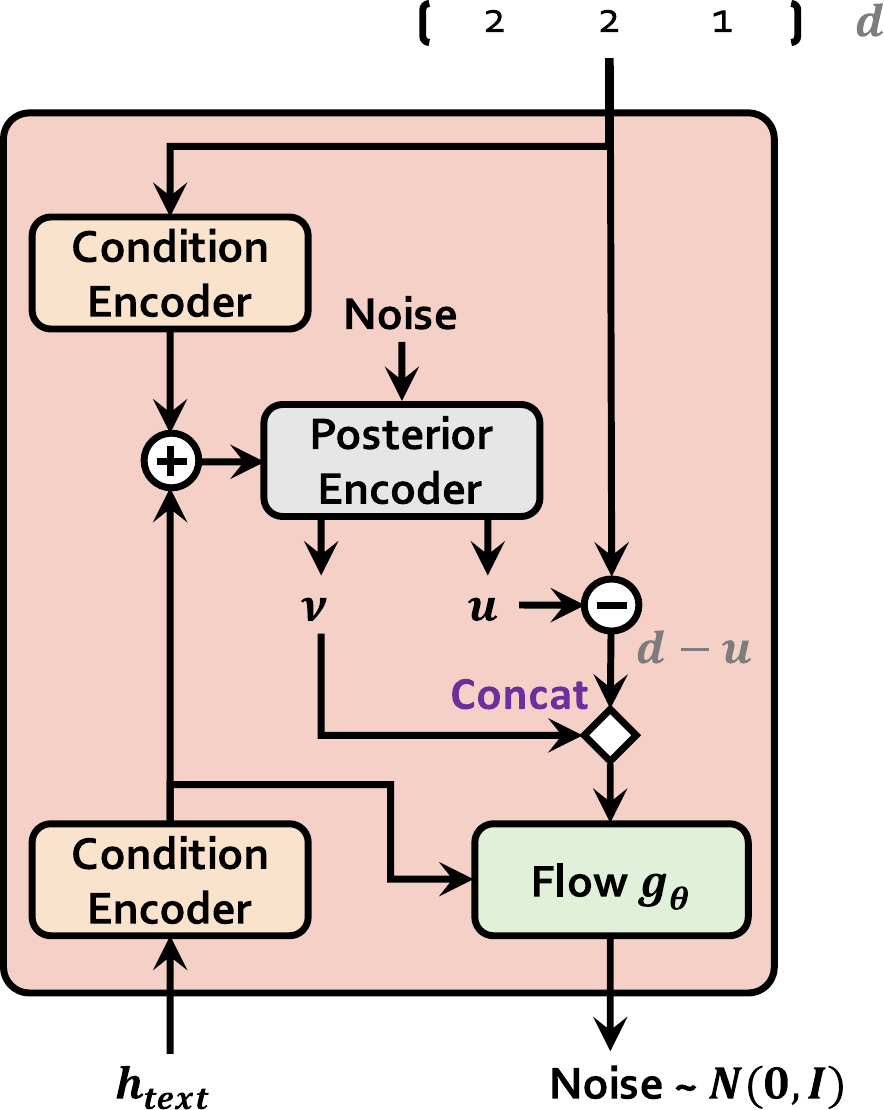}
            \vskip 0.1in
            \caption{Training procedure}
            \label{fig:sdp_training}
        \end{subfigure}\hfill%
        \begin{subfigure}{.32\textwidth}
            \centering
            \includegraphics[width=1.\linewidth]{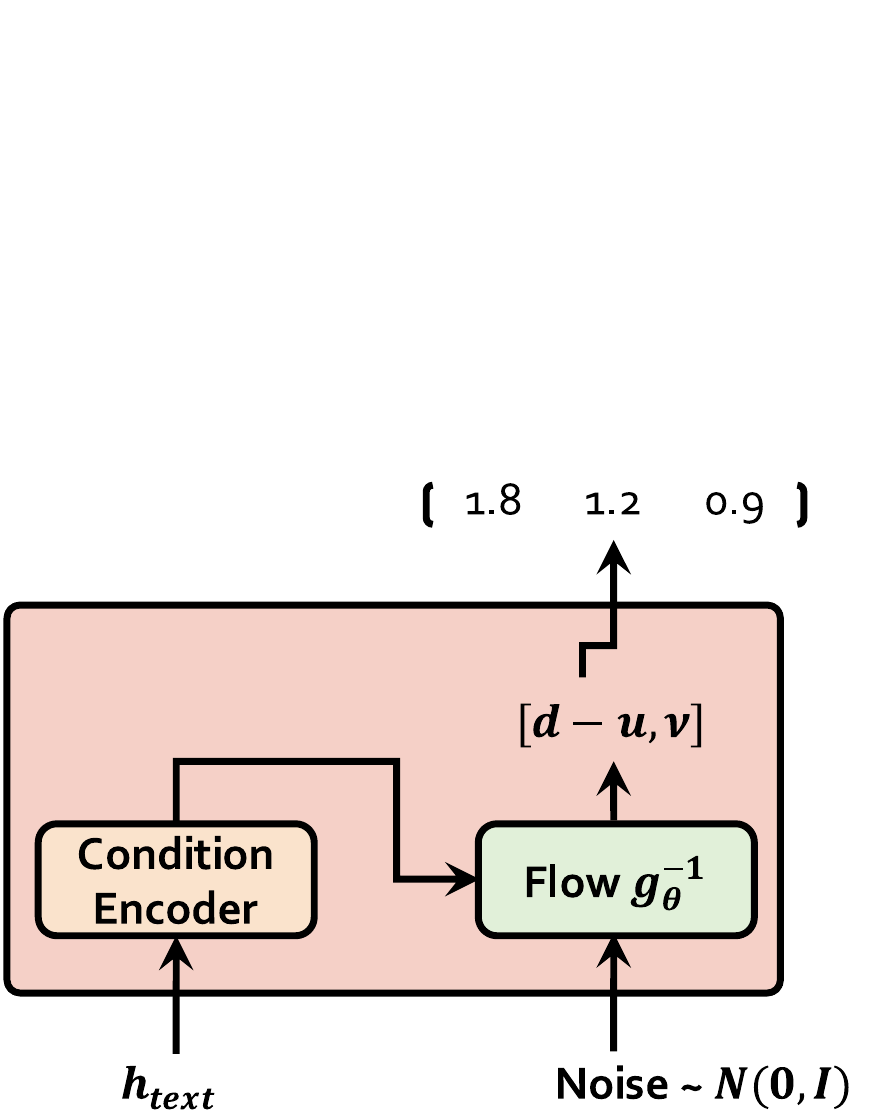}
            \vskip 0.1in
            \caption{Inference procedure}
            \label{fig:sdp_sampling}
        \end{subfigure}\hfill%
        \begin{subfigure}{.32\textwidth}
            \centering
            \includegraphics[width=1.\linewidth]{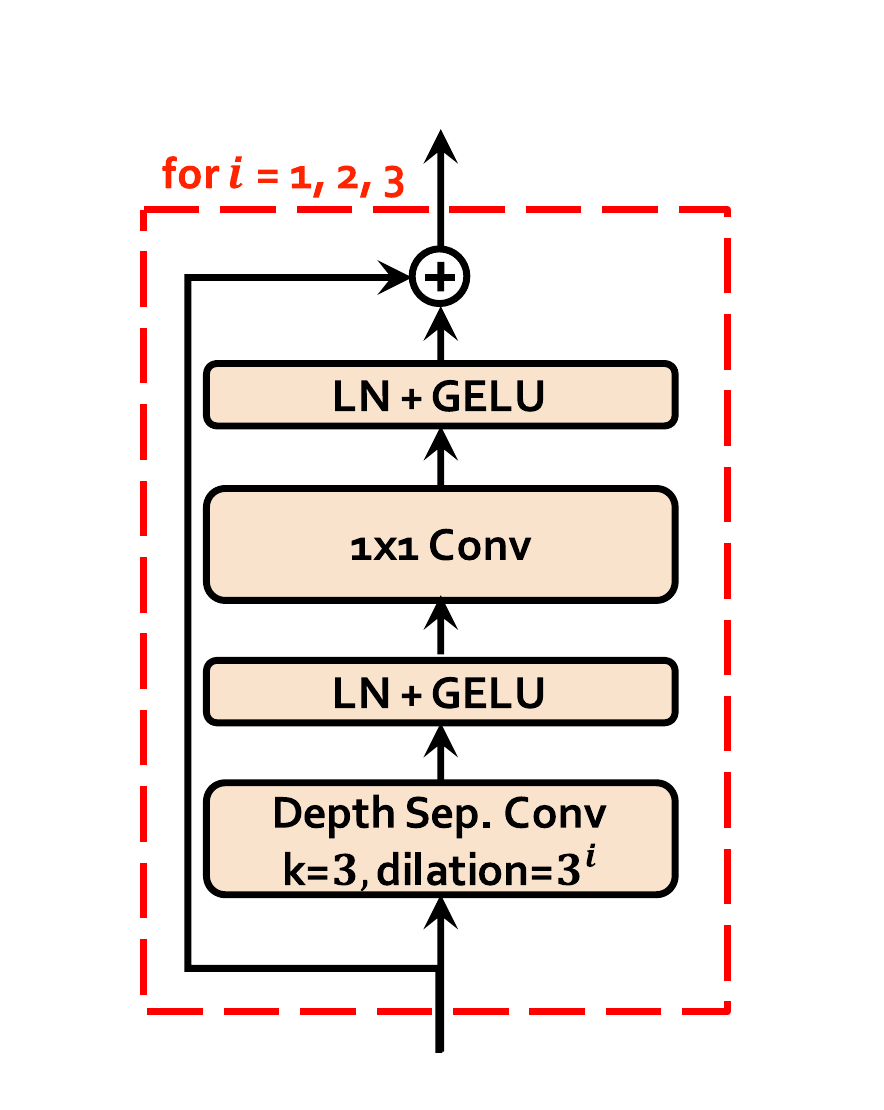}
            \vskip 0.1in
            \caption{Dilated and depth-wise separable convolutional residual block}
            \label{fig:sdp_dds}
        \end{subfigure}\hfill%
    \end{center}
    \vskip -0.05in
    \caption{Block diagram depicting (a) training procedure and (b) inference procedure of the stochastic duration predictor. The main building block of the stochastic duration predictor is (c) the dilated and depth-wise separable convolutional residual block.}
\end{figure*}

\subsection*{B.3. Stochastic Duration Predictor} Figures~\ref{fig:sdp_training} and \ref{fig:sdp_sampling} show the training and inference procedures of the stochastic duration predictor, respectively. The main building block of the stochastic duration predictor is the dilated and depth-wise separable convolutional (DDSConv) residual block as in Figure~\ref{fig:sdp_dds}. Each convolutional layer in DDSConv blocks is followed by a layer normalization layer and GELU activation function. We choose to use dilated and depth-wise separable convolutional layers for improving parameter efficiency while maintaining large receptive field size.

The posterior encoder and normalizing flow module in the duration predictor are flow-based neural networks and have the similar architecture. The difference is that the posterior encoder transforms a Gaussian noise sequence into two random variables $\nu$ and $u$ to express the approximate posterior distribution $q_{\phi}(u,\nu|d,c_{text})$, and the normalizing flow module transforms $d-u$ and $\nu$ into a Gaussian noise sequence to express the log-likelihood of the augmented and dequantized data $\log{p_{\theta}(d-u,\nu|c_{text})}$ as described in Section~\hyperref[sec/dp]{2.2.2}. 

\begin{figure*}[h]
    \vskip 0.1in
    \begin{center}
        \begin{subfigure}{.49\textwidth}
            \centering
            \includegraphics[width=.5\linewidth]{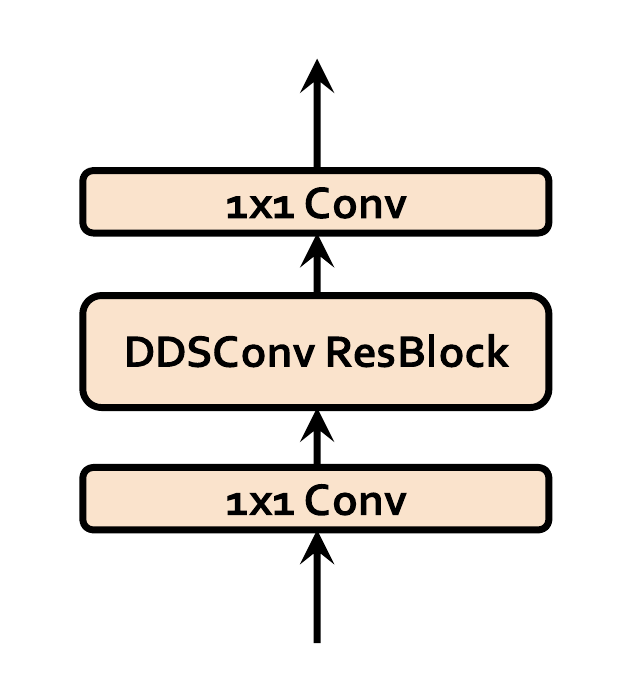}
            \vskip 0.1in
            \caption{Condition encoder in the stochastic duration predictor}
            \label{fig:sdp_cond}
        \end{subfigure}\hfill%
        \begin{subfigure}{.49\textwidth}
            \centering
            \includegraphics[width=.5\linewidth]{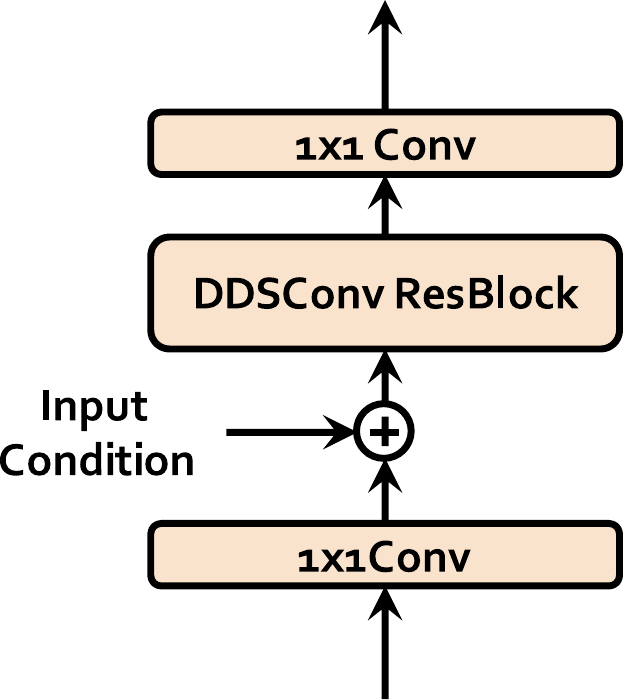}
            \vskip 0.1in
            \caption{Coupling layer in the stochastic duration predictor}
            \label{fig:sdp_flow}
        \end{subfigure}\hfill%
    \end{center}
    \vskip -0.05in
    \caption{The architecture of (a) condition encoder and (b) coupling layer used in the stochastic duration predictor.}
\end{figure*}

All input conditions are processed through condition encoders, each consisting of two 1x1 convolutional layers and a DDSConv residual block. The posterior encoder and normalizing flow module have four coupling layers of neural spline flows. Each coupling layer first processes input and input conditions through a DDSConv block and produces 29-channel parameters that are used to construct 10 rational-quadratic functions. We set the hidden dimension of all coupling layers and condition encoders to 192. Figure~\ref{fig:sdp_cond} and \ref{fig:sdp_flow} show the architecture of a condition encoder and a coupling layer used in the stochastic duration predictor.

\section*{C. Side-by-Side Evaluation }
\label{appb2}
We conducted 7-point Comparative Mean Opinion Score (CMOS) evaluation between VITS and the ground truth through 500 ratings on 50 items. Our model achieved -0.106 and -0.270 CMOS on the LJ Speech and the VCTK datasets, respectively, as in Table~\ref{tab:cmos}. It indicates that even though our model outperforms the best publicly available TTS system, Glow-TTS and HiFi-GAN, and achieves a comparable score to ground truth in MOS evaluation, there remains a small preference of raters towards the ground truth over our model.

\begin{table}[h]
    \centering
    \caption{Evaluated CMOS of VITS compared to the ground truth.}
    \label{tab:cmos}
    \begin{tabular}{lc}
        \toprule
        \textbf{Dataset} & \textbf{CMOS}\\
        \midrule
        LJ Speech & -0.106 \\
        VCTK & -0.262 \\
        \bottomrule
    \end{tabular}
\end{table}

\section*{D. Voice Conversion}
In the multi-speaker setting, we do not provide speaker identities into the text encoder, which makes the latent variables estimated from the text encoder learn speaker-independent representations. Using the speaker-independent representations, we can transform an audio recording of one speaker into a voice of another speaker. For a given speaker identity $s$ and an utterance of the speaker, we can attain a linear spectrogram $x_{lin}$ from the corresponding utterance audio. We can transform $x_{lin}$ into a speaker-independent representation $e$ through the posterior encoder and the normalizing flow in the prior encoder:
\begin{align}
z &\sim q_{\phi}(z|x_{lin}, s)\\
e &= f_{\theta}(z|s)
\end{align}
Then, we can synthesize a voice $\hat{y}$ of a target speaker identity $\hat{s}$ from the representation $e$ through the inverse transformation of the normalizing flow $f^{-1}_{\theta}$ and decoder $G$:
\begin{align}
\hat{y} = G(f_{\theta}^{-1}(e|\hat{s})|\hat{s})
\end{align}

\begin{figure*}[h]
    \vskip 0.1in
    \centering
    \includegraphics[width=.6\linewidth]{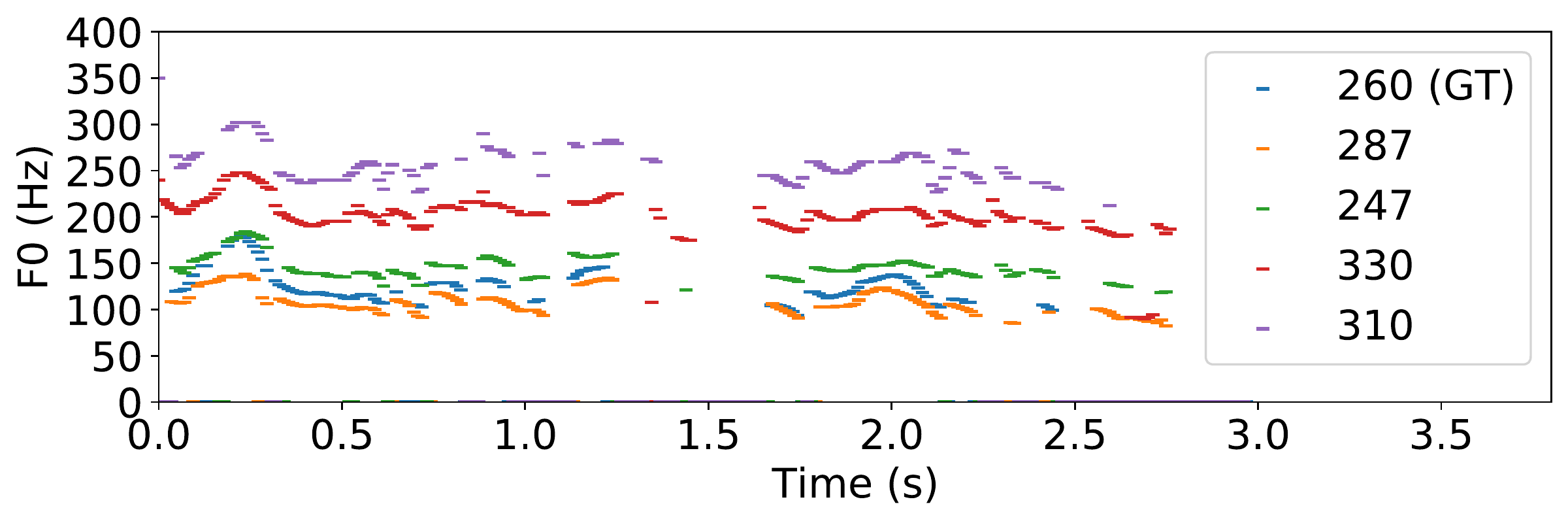}
    \vskip -0.05in
    \caption{Pitch tracks of a ground truth sample and the corresponding voice conversion samples with different speaker identities.}
    \label{fig:f0_vc}
\end{figure*}

Learning speaker-independent representations and using it for voice conversion can be seen as an extension of the voice conversion method proposed in Glow-TTS. Our voice conversion method provides raw waveforms rather than mel-spectrograms as in Glow-TTS. The voice conversion results are presented in Figure~\ref{fig:f0_vc}. It shows a similar trend of pitch tracks with different pitch levels.

%% file: figures/mas_pseudocode.tex
\begin{minted}{python}
def monotonic_alignment_search(value):
    """Returns the most likely alignment for the given log-likelihood matrix.
    Args:
        value: the log-likelihood matrix. Its (i, j)-th entry contains 
        the log-likelihood of the j-th latent variable
        for the given i-th prior mean and variance:
        .. math::
            value_{i,j} = log N(f(z)_{j}; \mu_{i}, \sigma_{i})
        (dtype=float, shape=[text_length, latent_variable_length])
    Returns:
        path: the most likely alignment.
        (dtype=float, shape=[text_length, latent_variable_length])
    """
    t_x, t_y = value.shape # [text_length, letent_variable_length]
    path = zeros([t_x, t_y])
    
    # A cache to store the log-likelihood for the most likely alignment so far.
    Q = -INFINITY * ones([t_x, t_y])
    
    for y in range(t_y):
        for x in range(max(0, t_x + y - t_y), min(t_x, y + 1)):
            if y == 0: # Base case. If y is 0, the possible x value is only 0.
                Q[x, 0] = value[x, 0]
            else:
                if x == 0:
                    v_prev = -INFINITY
                else:
                    v_prev = Q[x-1, y-1]
                v_cur = Q[x, y-1]
                Q[x, y] = value[x, y] + max(v_prev, v_cur)

    # Backtrack from last observation.
    index = t_x - 1
    for y in range(t_y - 1, -1, -1):
        path[index, y] = 1
        if index != 0 and (index == y or Q[index, y-1] < Q[index-1, y-1]):
            index = index - 1
    
    return path
\end{minted}